\documentclass[aps,prb,twocolumn,showpacs,amsmath,amssymb,superscriptaddress]{revtex4-1}
\usepackage{graphicx}% Include figure files
\usepackage{amssymb}
\usepackage{dcolumn}% Align table columns on decimal point
\usepackage{bm}% bold math
\usepackage{epstopdf}
\usepackage{float,epsfig}
\begin{document}

\title{Magnetic control of the valley degree of freedom of massive Dirac fermions
with application to transition metal dichalcogenides}

\author{Tianyi Cai}
\affiliation{School of Physical Science and Technology, Soochow University, Suzhou 215006, China}

\author{Shengyuan A. Yang}
\email{shengyuan\_yang@sutd.edu.sg}
\affiliation{Engineering Product Development, Singapore University of Technology and Design, Singapore 138682, Singapore}

\author{Xiao Li}
\affiliation{Department of Physics, The University of Texas at Austin, Austin, Texas 78712, USA}

\author{Fan Zhang}
\affiliation{Department of Physics and Astronomy, University of Pennsylvania, Philadelphia, Pennsylvania 19104, USA}

\author{Junren Shi}
\affiliation{International Center for Quantum Materials, Peking University, Beijing 100871, China}

\author{Wang Yao}
\affiliation{Department of Physics and Center for Theoretical and Computational Physics, The University of Hong Kong, Hong Kong}

\author{Qian Niu}
\affiliation{Department of Physics, The University of Texas at Austin, Austin, Texas 78712, USA}
\affiliation{International Center for Quantum Materials, Peking University, Beijing 100871, China}

\begin{abstract}
We study the valley-dependent magnetic and transport properties of massive
Dirac fermions in multivalley systems such as the transition metal dichalcogenides.
The asymmetry of the zeroth Landau level between valleys and the
enhanced magnetic susceptibility can be attributed to the different orbital
magnetic moment tied with each valley. This allows the valley polarization to be controlled by
tuning the external magnetic field and the doping level.
As a result of this magnetic field induced valley polarization,
there exists an extra contribution to the ordinary Hall effect. All these
effects can be captured by a low energy effective theory with a valley-orbit coupling term.

\end{abstract}

 \maketitle

\section{Introduction}

Many crystalline materials have multiple energy extremum points near the Fermi level
in the reciprocal space, which are related to each other through symmetry operations
and are referred to as valleys. A well-known example is Si, which has six
valleys at its conduction band edge in the Brillouin zone. Similar to spin, the valley
labeling constitutes a discrete degree of freedom. Therefore, it was proposed that
the valley degree of freedom could be used for the information coding and transmission,
giving rise to an active research field called valleytronics.\cite{ryce2007,xiao2007,akhm2007,yao2008,yao2009,zhang2011,zhu2012}

To have a successful valleytronics application, there are at least two basic requirements.
The first is the ability to generate and control the valley polarization, and the second
is the ability to detect the valley polarization. For conventional semiconductor materials
such as Si, it is difficult to distinguish different valleys. The situation changes with the
advent of novel two dimensional (2D) materials which support massive Dirac fermion excitations.
Examples include graphene\cite{novo2004,novo2005,zhan2005} with sublattice symmetry breaking,
silicene, transition metal
dichalcogenides \emph{etc.}\cite{novo2005a} In these systems, there are two inequivalent valleys K and K'
located at the corners of the hexagonal Brillouin zone. The special feature of the massive
Dirac fermion type excitation is that each valley has a definite chirality arising from
its strong pseudospin-orbit coupling.\cite{xiao2007} More importantly, the chirality of the two valleys
are opposite to each other which is imposed by the time-reversal and inversion symmetries. This leads
to possible practical ways to differentiate the two valleys and to address them individually.\cite{xiao2007,yao2008}

Of the examples mentioned above, 2D transition metal dichalcogenides (TMD) is especially interesting
and has attracted a lot of attention recently.\cite{radi2011,wang2012,sple2010,mak2010,korn2011,xiao2012,zeng2012,mak2012,cao2012,wu2013,jone2013,mak2013,ross2013,gong2013,ye2013,Bolotin} It has been found that when thinned down to a single
layer, several members of this class of materials undergo a transition from indirect band gap
to direct band gap with a gap size of $1\sim 2$eV, which is suitable for optical manipulations.\cite{sple2010,mak2010,korn2011}
It has been successfully demonstrated that the excitonic valley polarization and coherence in 2D TMD can be generated by pumping with circularly polarized light and linearly polarized light respectively.\cite{xiao2012,zeng2012,mak2012,cao2012,wu2013,jone2013} The optically generated excitonic states could
be manipulated electrically,\cite{mak2013,ross2013} and have a long spin coherence time due to the large valley separation and the large spin splitting in the valence band.\cite{xiao2012,jone2013} The field effect transistors with a single
layer of MoS$_2$ has also been fabricated and the mobility can be
enhanced to $~500$cm$^2$/(V$\cdot$s)\cite{radi2011,ye2013,Bolotin} with an excellent current on/off ratio.

Motivated by these recent progresses, we explore the possibility of controlling the
valley degree of freedom in TMD through magnetic means. We show that the zeroth Landau level
anomaly which was found previously in MoS$_2$\cite{li2013} and the enhanced magnetic susceptibility
can be attributed to the valley-contrasting orbital magnetic moments. With this property, we
could generate valley polarization of carriers by using an external magnetic field. Furthermore,
this induced valley polarization would in turn produce an extra contribution to the ordinary
Hall effect which can be detected experimentally. A simple effective theory is
proposed to describe the dynamics of such valley-orbit coupled systems. These findings may open
a new route for the valleytronics applications.

This paper is organized as follows. In section II we
discuss the simplest model of massive Dirac fermions, which serves
as the generic building blocks for the more realistic models. In section III, we
apply the results from section II to study the TMD materials
and discuss its Landau level structures and magnetic susceptibility. In section IV
we show that
the magnetic field can be used to control the valley polarization.
In sections V we predict that this valley polarization leads to an extra contribution
in the charge Hall transport.
Some discussions and a summary is presented in section VI.

\section{simple model of massive Dirac fermions}

In this section, we first present a heuristic discussion of the
simplest model for a massive Dirac fermion with valley degrees of freedom.
In the absence of external fields, the model can be written as
\begin{equation}\label{H0}
H_\text{D}=\hbar v(\tau_z k_x \sigma_x+k_y \sigma_y)+\Delta\sigma_z,
\end{equation}
where $v$ is a material specific Fermi velocity,
$k_x$ and $k_y$ are the two components of the wave vector measured from the Dirac point,
$\sigma$'s are the Pauli matrices typically representing a pseudospin
from the sublattice or orbital degrees of freedom, and $\tau_z=\pm 1$ is the
valley index labeling the two inequivalent
valleys. The form of the Hamiltonian is generic for several 2D materials we are interested in
including the TMD materials. These materials usually have a honeycomb lattice structure
when viewed from the top. The two valleys occur at K and K' points at the corners of the
hexagonal Brillouin zone, and are related to each other
by the time-reversal and inversion symmetries.

The first term in $H_0$ shows a strong pseudospin-orbit coupling
and the second term is the mass term.
If $\Delta\rightarrow 0$, the particle becomes massless and the Hamiltonian can be used to
describe the graphene. For a finite $\Delta$, it opens a gap of $2\Delta$ in the spectrum.

The two valleys have different chiralities. This can be understood by tracking the pseudospin
orientation when an electron moves around a fixed energy contour
enclosing a Dirac point.
Consider an electron going around the states
(counterclockwise) with a fixed energy $\varepsilon> 2\Delta$ and returning to its starting point.
Its pseudospin $\bm \sigma$ would rotate by $+2\pi$ for the K valley ($\tau_z=+1$), while it would rotate by $-2\pi$
for the K' valley ($\tau_z=-1$).
This difference in the chirality or the winding numbers manifests in many important electronic properties
such as the Berry curvature and the orbital magnetic moment.\cite{xiao2007,zhang2011}

\subsection{Asymmetric Landau Levels}
The effect of external magnetic field (oriented
in the $z$-direction, i.e. perpendicular to the plane) can be taken into account through
the Peierls substitution of $\bm k$ by
$\bm \pi=\bm k +e\bm A/\hbar$ where $\bm A$ is the vector potential.
Here we neglect the Zeeman energy term which is at least one order
of magnitude smaller than the cyclotron energy.\cite{li2013} Hence in the remaining part of this section
we shall neglect the spin degeneracy.
Following the standard procedure for the Landau level quantization, we define the
operators $\pi_{\pm}=\pi_{x}\pm
i\pi_{y}$ which satisfy the commutation relation
$[\pi_{-},\pi_{+}]=2eB/\hbar$ where $B$ is the magnitude of the
magnetic field. Hence we could define the bosonic ladder
operators $b^\dagger$ and $b$ as $b^\dagger=(l_B/\sqrt{2}) \pi_+$,
$b=(l_B/\sqrt{2}) \pi_-$, where $l_B=\sqrt{\hbar/(eB)}$ is the magnetic length.
These ladder operators
satisfy the relations $b|n\rangle=\sqrt{n}|n-1\rangle$, $b|0\rangle=0$, where $|n\rangle$ ($n=0,1,2,\cdots$)
are the Landau level eigenstates for a conventional 2D electron gas.

The spectrum can be easily solved in the basis of $|n\rangle$'s. The resulting
Landau levels are
\begin{equation}\label{llevel}
\varepsilon _{n,\pm} = \tau_z\Delta\delta_{n,0}\pm \sqrt{\Delta^2+n\hbar^2\omega_c^2}(1-\delta_{n,0}),
\end{equation}
where $\omega_c=\sqrt{2}v/ l_B$, $\delta$ is the Kronecker delta function, and $n$ is an integer
$\geq 0$.

\begin{figure}[!]
\includegraphics[scale=0.6]{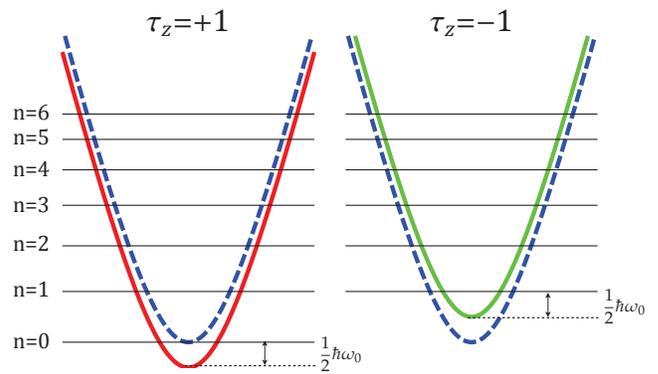}
\caption{(Color online). The Landau levels from exact quantum
calculations for the conduction band of the two valleys, shown as solid horizontal lines.
The original band dispersion
in the absence of fields is shown as dashed blue curves.
The solid curves show the band dispersion shifted by
the Zeeman-like coupling between magnetic moment $\bm m(\bm k)$ and
the magnetic field. It can be seen that the first Landau level
is at $\hbar\omega_0/2$ above the bottom of the shifted band,
where $\hbar\omega_0$ is the cyclotron energy of the first Landau level.\label{Fig:LandauLevels}}
\end{figure}

We observe that the Landau level spacing is not uniform (see Fig.~\ref{Fig:LandauLevels}).
The Landau levels with $n\geq 1$ are aligned between
the two valleys.
However, the zeroth Landau level with $n=0$ is not located at
the zero energy,
and for different valleys its position shifts in opposite directions.
Note that the zeroth Landau level for $\tau_z=+1$ valley is at the same energy of
the original conduction band bottom at zero field, while for the other valley
there is no Landau level at this energy. The spacing between the zeroth
and the first Landau level in $\tau_z=+1$ valley is
\begin{equation}\label{LLS}
\delta\varepsilon=\sqrt{\Delta^2+\hbar^2\omega_c^2}-\Delta \approx
\frac{e\hbar v^2}{\Delta}B,
\end{equation}
where in the second step we assume the gap is much larger than the cyclotron energy.

This peculiar asymmetric behavior can be traced to the chirality difference
between the two valleys and can be easily explained in the semiclassical theory
of Bloch electrons. If we construct an electron wave-packet
near the valley center,
due to the pseudospin-orbit coupling, it is self-rotating hence
producing an intrinsic orbital magnetic moment.
The general expression is given by\cite{chan1996,sund2000,xiao2010}
\begin{equation}\label{littlem}
\bm m(\bm k)=-i\frac{e}{2\hbar}\langle\nabla_{\bm k}u|\times
[H(\bm k)-\varepsilon(\bm k)]|\nabla_{\bm k}u\rangle,
\end{equation}
where $|u\rangle$ is the periodic part of the Bloch eigenstate, $H(\bm k)$ is the Bloch Hamiltonian
and $\varepsilon(\bm k)$ is the band energy. Generally speaking,
the orbital magnetic moment is large for states near gaps caused by (pseudo-)spin-orbit coupling.
For our present case, a direct calculation yields
\begin{equation}
\bm{m}(\bm{k}) =\tau _z
\frac{{ e\hbar v^2\Delta  }}{{2 \left( {\Delta ^2  +\hbar^2 v^2 k^2 } \right)}}\hat{\bm z}.
\end{equation}
This moment is largest
at the band edges $\pm \Delta$. For example, at conduction band bottom,
it is given by
\begin{equation}\label{moment}
\bm m=\tau_z\frac{e\hbar}{2m_e^*}\hat{\bm z},\qquad\text{with}\;m_e^*=\frac{{ \Delta }}{{v^2 }}.
\end{equation}
Here we express the moment in a form similar to that for the Bohr magneton. The
difference is that now the bare electron mass is replaced by an effective mass
$m_e^*$ determined by the band parameters. It is noted that the magnitude of this moment
is proportional to $\Delta^{-1}$, i.e. the moment becomes larger when the gap is smaller.
The most important feature for our purpose is that it takes different signs
for the two valleys. This is a manifestation of the valley-dependent chirality
of the two valleys we mentioned before,
because chirality can be viewed as
representing a sense of rotation of the carrier.
It is an important property of the electronic band structure.

In the presence of an external magnetic field, the wave-packet
energy would be shifted by $-\bm m\cdot \bm B$.\cite{chan1996} We plot the shifted bands in
Fig.~\ref{Fig:LandauLevels}. The bands at the two valleys are shifted in opposite direction, in particular
the band edges are shifted by
\begin{equation}
\varepsilon_m=-\tau_z\frac{e\hbar v^2}{2\Delta}B.
\end{equation}
Its magnitude is equal to $\delta\varepsilon/2$, i.e. half the spacing between
the zeroth and the first Landau levels.
Now we can see that with respect to the shifted bands, the Landau levels at both valleys
start at half of the cyclotron energy above the band edges,
which is just the familiar result for the 2D free electron gas. This implies that
the behavior of the present system under a magnetic field can be understood
using a 2D electron gas with an valley-contrasting intrinsic magnetic moment.
This point of view would be further supported in the next subsection
when we study the magnetic susceptibility. In fact all the Landau level spectrum
could be obtained in the semiclassical theory through the semiclassical quantization
procedure as outlined in the appendix.

\subsection{Enhanced Magnetic Susceptibility}

\begin{figure}[!]
\includegraphics[scale=0.9]{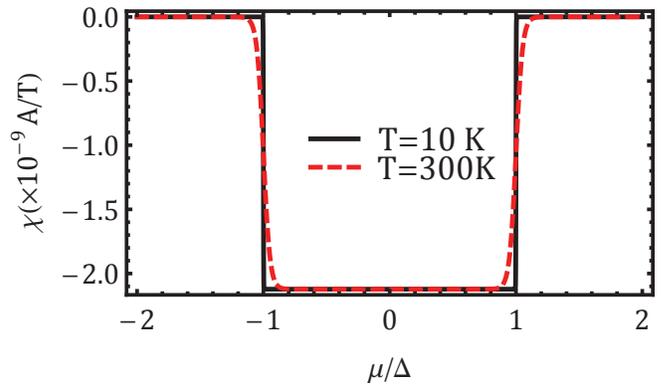}
\caption{(Color online). Magnetic susceptibility as a function
of the chemical potential $\mu$. Two different temperatures of $T=10$\,K and $T=300$\,K are taken.
Here we take $v=0.5\times 10^6$\,m/s.\label{Fig:MagneticSusceptibillity}}
\end{figure}

The magnetic susceptibility captures the collective response of the system
to the external magnetic field. It can be calculated from the thermodynamic potential
\begin{equation}
F = -\frac{1}{\beta}\frac{{eB}}{h}\text{Tr} \left\{{\ln [ {1 +
e^{\beta (\mu - \hat{H} )} } ]}\right\},
\end{equation}
where $\beta=1/(k_B T)$ is the inverse temperature and $h$ is the Planck constant. The magnetic susceptibility
can be extracted as  $ \chi=  - \left(
\partial^2 F/\partial B^2 \right)_{\mu,B\rightarrow 0}$.
Substituting in the Landau level spectrum, we can expand $F$ as a power series
in the field strength $B$ and obtain an analytical expression of susceptibility:\cite{shar2004}
\begin{equation}\label{chi0}
\chi_0(\mu;\Delta)=-\frac{e^2v^2}{6\pi}\frac{\sinh(\beta\Delta)}{\Delta[\cosh(\beta\mu)+\cosh(\beta\Delta)]}.
\end{equation}
In the limit of zero gap $\Delta\rightarrow 0$, we have
\begin{equation}
\lim_{\Delta\rightarrow 0}\chi_0=-\frac{e^2v^2}{6\pi}\frac{\beta}{1+\cosh(\beta\mu)},
\end{equation}
which recovers the old result discussed in the context of graphene.\cite{mccl1956,ando2007} We observe that
the $\chi_0$ in the zero gap case has a large negative peak at $\mu=0$, and it
diverges as $T\rightarrow 0$. The divergence is removed when the gap opens up,
but the large diamagnetic dip is still visible and it gets broadened in energy as the
gap increases (see Fig.~\ref{Fig:MagneticSusceptibillity}). It is interesting to note that the integral of susceptibility over chemical
potential,
\begin{equation}
\int_{-\infty}^{+\infty} {\chi_0 (\mu )\text{d}\mu }=-\frac{e^2v^2}{3\pi} ,
\end{equation}
which is independent of both
the gap size and the temperature. At zero temperature limit, the susceptibility
becomes a square well shape, and completely vanishes outside the gap.\cite{note1}

The sudden jump of magnetic susceptibility at band edges in fact signifies a
large paramagnetic response from the carriers. Indeed, if we calculate the
magnetic susceptibility from the Landau levels above the gap, then the contribution
from the conduction band carriers can be obtained as
\begin{equation}\label{chiC}
\chi _0^c(\mu)  = \frac{{e^2 v^2 }}{{6\pi  }}\frac{1}{\Delta}
\frac{1}{{1 + e^{ - \beta (\mu  - \Delta)} }},
\end{equation}
which shows a large paramagnetic
response from the conduction band electrons.

The large paramagnetic response at the band edge can also be understood in our semiclassical picture
mentioned before. At the conduction band bottom the system can be viewed as
a 2D gas of electron wave-packets with magnetic moments $\bm m$.
In the presence of an external magnetic field, the resulting magnetic response
is well-known. It has a net paramagnetic response
from the difference between the Pauli paramagnetism and the Landau diamagnetism~\cite{mard2010}.
We have checked that the result just recovers Eq.~\eqref{chiC} above. Similarly, the
large diamagnetic response in the gap can be attributed to the orbital
magnetic moments concentrated at the valence band top. In the zero gap limit,
$m^*_e\rightarrow 0$ and the moment diverges at the band edges, which results in
the singular behavior of the magnetic susceptibility.\cite{mccl1956}

\section{magnetic response of transition metal dichalcogenides}

In this section, we apply the knowledge obtained from the last section to study
the magnetic response of a monolayer TMD material. Physically interesting TMD materials have the
form of MX$_2$ (M=Mo, W; X=S, Se).\cite{wang2012} As we mentioned earlier, these materials have
direct band gaps at K and K' points. They also have a large spin-orbit coupling induced
spin splitting in the valence band. The electronic properties near the band edges
can be described by the effective Hamiltonian\cite{xiao2012}
\begin{equation}\label{H}
H=\hbar v(\tau_z k_x \sigma_x+k_y \sigma_y)+\Delta\sigma_z-\lambda\tau_zs_z\sigma_z+\lambda\tau_zs_z.
\end{equation}
Note that the first two terms are the same as those in our simple massive
Dirac model Eq.~\eqref{H0}. Now the pseudospin $\sigma$ represents the space
of two relevant orbitals $d_{z^2}$ and $d_{x^2-y^2}+id_{xy}$. The last two terms
represent the effect of spin-orbital
coupling induced spin splitting. It splits the valence band top
into two spin polarized bands. $\lambda$ is the coupling strength
and $s_z=\pm 1$ is for the $z$-component of real spin.

\begin{figure}
\includegraphics[scale=0.6]{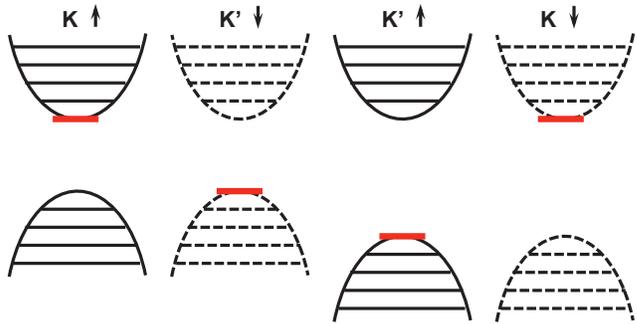}
\caption{Landau levels for the massive Dirac fermions in transition metal dichalcogenides.
Solid (dashed) curves represent spin up (down) bands, while the parallel lines represent their Landau levels. The four red lines represent the location of the n=0 Landau levels. \label{Fig:zeroLLTMD}}
\end{figure}

We note that because of the extra spin degrees of freedom, the model in fact
consists of two copies of the simple model (\ref{H0}) with different band gaps (masses)
depending on the ``flavor" index $\tau_zs_z$.
The gap is $2(\Delta-\lambda)$ for $\tau_zs_z=+1$, and is $2(\Delta+\lambda)$ for $\tau_zs_z=-1$.
In this way the spin and valley are coupled together. The valence band top in K valley
is polarized with spin up while in the K' valley it is polarized with spin down.
Accordingly there are two sets of Landau levels which can be written in a compact way as\cite{li2013}
\begin{equation}
\varepsilon_{n,\pm}=\lambda\tau_zs_z\pm\sqrt{\tilde{\Delta}^2+n\hbar^2\omega_c^2},
\end{equation}
where $\tilde{\Delta}(\tau_z,s_z)=\Delta-\lambda\tau_zs_z$ is the flavor dependent mass and
$n=0,1,2\cdots$ is a non-negative integer. Again the asymmetric Landau
level structure between the two valleys can be observed. It is also interesting
to note that the Landau levels near the valence band top at K' valley is spin
polarized (see Fig.~\ref{Fig:zeroLLTMD}).

\begin{figure}
\includegraphics[scale=0.9]{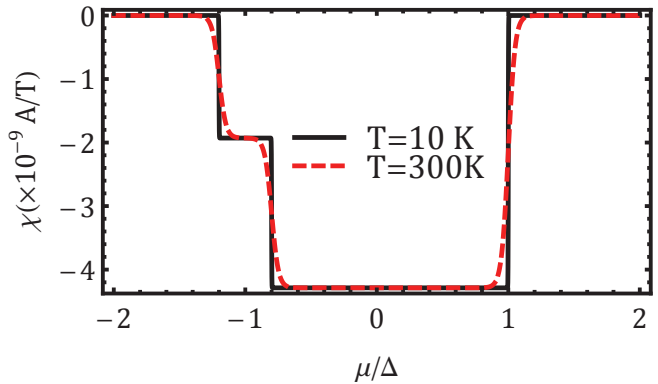}
\caption{(Color online). Magnetic susceptibility of typical TMD materials. Two different temperatures of $T=10$\,K and $T=300$\,K are taken. Here we take the spin-orbit coupling strength to be $\lambda=0.1\Delta$,
and $v=0.5\times 10^6$\,m/s. \label{Fig:MagneticSusceptibillityTMD}}
\end{figure}

The magnetic susceptibility can be directly read out by combining the contributions from the
two copies of massive Dirac bands:
\begin{equation}
\chi(\mu)=\chi_0(\mu-\lambda;\Delta-\lambda)+\chi_0(\mu+\lambda;\Delta+\lambda),
\end{equation}
where $\chi_0(x;y)$ is defined in Eq.~\eqref{chi0}. Due to the orbital magnetic moments
of carriers at the band edges, again a sudden change of orbital magnetic susceptibility
is expected there. This behavior is shown in Fig.~\ref{Fig:MagneticSusceptibillityTMD}. And because here we have two copies of
massive Dirac bands with different gap sizes, there appears a two-step feature in the
susceptibility curve plotted as a function of the chemical potential. The steep change
feature is going to be smoothed out as temperature increases.

\section{magnetic field induced valley polarization}

A necessary requirement for valleytronics applications is the ability to generate and control
the valley polarization. It has been shown that the valley polarization in TMD
can be generated by a circularly polarized light which couples to the orbital magnetic moment.
Here we show that due to the valley-contrasting orbital magnetic moment,
we can also control the valley polarization in TMD by an applied magnetic field.

First analogous to the spin polarization, we
can define the valley polarization of carriers as
\begin{equation}
P_\text{v}=\frac{n_{ + } - n_{ - }}{n },
\end{equation}
where $n_{ \pm}$ is the density of electrons in the
valley with index $\tau_z=\pm 1$, and $n=n_{ + } + n_{
- } $ is the total carrier density. In an external magnetic field, $n_\pm$
can be calculated from the
filling of the Landau levels in the two valleys. Now it is easy to see that
a finite valley polarization must be present due to the asymmetric Landau
level structure of the two valleys.

Considering the n-doped case, because the Landau levels (there are two levels from the
spin degeneracy) at $+\Delta$ for K valley have no counterpart at K' valley,
hence for low doping such that only these two lowest levels are filled,
we can achieve 100\% valley polarization. If we keep on increasing the doping,
higher Landau levels from both valleys are going to be filled. The valley polarization
decreases and approaches zero as $n\rightarrow\infty$. The situation is similar for the p-doped case.
More interestingly, because of the spin splitting at the valence band top,
the valley polarization is also the spin polarization. Specifically, at a low doping level,
only the Landau level at $-\Delta+2\lambda$ at K' valley is occupied by
holes with spin up. Moreover, for both cases, the valley polarization
can be reversed by simply reversing the direction of the magnetic field.

Let's take the simple $T=0$ limit. Then the polarization can be
obtained by simply counting the Landau levels.
For the n-doped case, the variation of $P_\text{v}$
as a function of the conduction band electron density $n_e$ is given by
(note that there are two spin degenerate zeroth levels at K valley)
\begin{equation}
P_{\text{v}}=\Theta(2-\nu_e)+\frac{2}{\nu_e }\Theta(\nu_e-2),
\end{equation}
where $\Theta$ is the step function, and $\nu_e=h n_e/(eB)$
counts the filling of Landau levels. Similarly for the weakly p-doped case, i.e. when the Fermi level is above
the lower spin-split band,
\begin{equation}
P_{\text{v}}=-\Theta(1-\nu_h)-\frac{1}{\nu_h }\Theta(\nu_h-1).
\end{equation}
Here $\nu_h=h n_h/(eB)$ with $n_h$ being the hole density.

\begin{figure}[!]
\includegraphics[scale=0.9]{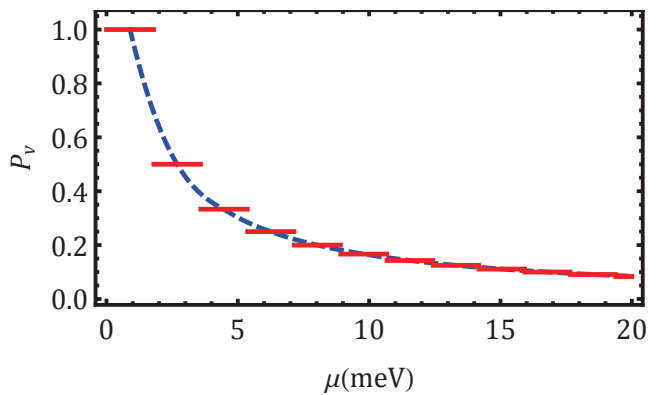}
\caption{(Color online). The variation
of valley polarization $P_\text{v}$ with the chemical potential $\mu$. Here we take $v=0.5\times 10^6$\,m/s, $\Delta=0.5$\,eV, the spin-orbit coupling strength $\lambda=0.05$\,eV, and the magnetic field $B=5$\,T. The red steps represent exact quantum calculations, while the blue dashed curve
represents the semiclassical results. The chemical potential is measured from the conduction band $n=0$ Landau level. \label{Fig:Polarization}}
\end{figure}

The valley polarization as a function
of the chemical potential would show a series of steps at low temperatures.
Again taking the $T=0$ limit, we have for the n-doped case
\begin{equation}\label{Pv}
P_\text{v}=\frac{1}{1+m}, \qquad E_m\leq \mu<E_{m+1}
\end{equation}
where $E_m$ is the $m$-th Landau level energy for the conduction band
including both flavors $\tau_z s_z=\pm 1$
and $E_0=+\Delta$ is the zeroth Landau level energy at the K valley.
The result is plotted in Fig.~\ref{Fig:Polarization}.
For the weakly p-doped case the valley polarization is given by
\begin{equation}
P_\text{v}=-\frac{1}{1+2m}, \qquad E_{m+1}< \mu\leq E_{m},
\end{equation}
where we count the Landau levels from $\mu=0$ with decreasing energy and
$E_0=-\Delta+\lambda$ is the energy of the zeroth level at the K' valley.
At finite temperature, the sharp changes of the above functional dependence
of $P_\text{v}$ are going to be smoothed out, but the main features should
be maintained.

The valley polarization can also be calculated using the semiclassical
theory, in which the electron density from a given valley is obtained by
integrating the Berry curvature modified density of states\cite{xiao2005}
\begin{equation}
n(\mu,\tau_z,s_z)=\int^\mu\frac{\text{d}^2k}{(2\pi)^2}\left(1+\frac{\bm  B\cdot\bm \Omega}{\hbar}\right)f(k,\tau_z,s_z),
\end{equation}
where the Berry curvature $\bm \Omega(\bm k)=i\left\langle \nabla_{\bm k}u|\times|\nabla_{\bm k}u\right\rangle$
is also an intrinsic band property like the orbital magnetic moment, and $f$ is the Fermi distribution function.
The effect of the external field is in the shift of the bands
$f=f[\varepsilon(k)-m(k)B]$ ($\varepsilon$ and $m$ also depends on $\tau_z$ and $s_z$).
The resulting valley polarization is a smooth curve going through all the steps of the exact quantum result
Eq.(\ref{Pv}) (see Fig.~\ref{Fig:Polarization}). It is a good approximation in the low field regime, as it should be from the condition of validity
of the semiclassical theory.

\section{Hall transport from valley polarization}

It is now clear that even in the absence of the external magnetic field,
transverse motion of the carriers could be induced due to the Berry curvature
which acts like an effective magnetic field in the reciprocal space.\cite{jung2002,onod2002} Like the orbital magnetic moment,
the Berry curvature is also related to the chirality of the band.
For the present model of TMD,
the broken inversion symmetry leads to a nontrivial momentum-space Berry curvature which reads\cite{xiao2012,li2013}
\begin{equation}
\Omega_c(k,\tau_z,s_z)=-\tau_z\frac{\hbar^2 v^2\tilde{\Delta}}{2(\tilde{\Delta}^2+\hbar^2 v^2k^2)^{3/2}}.
\end{equation}
The appearance of the factor of $\tau_z$ indicates it is also a valley contrasting property. For the valence
band the Berry curvature has the opposite sign $\Omega_v(k,\tau_z,s_z)=-\Omega_c(k,\tau_z,s_z)$.

In the presence of an in-plane electric field,
this Berry curvature leads to the transverse motion of the carriers.
Its integral over the occupied states gives an intrinsic contribution
to the Hall conductivity,\cite{jung2002,onod2002}
\begin{equation}
\sigma_\text{H}^\text{int}=\frac{e^2}{\hbar}\sum_{\tau_z,s_z} \int
\frac{\text{d}^2k}{(2\pi)^2}f(k,\tau_z,s_z)\Omega(k,\tau_z,s_z),
\end{equation}
where $f$ is the Fermi-Dirac distribution function. Disorder scattering also contributes
to the Hall transport. There is an important side-jump contribution which is
proportional to the Berry curvature at the Fermi surface.\cite{berg1970}
In the following, we assume the scattering is of Gaussian white noise type,\cite{note2}
and disregard the inter-valley scattering
which requires a large momentum transfer.

We first consider a single copy of massive Dirac fermion as in Eq.(\ref{H0}).
For the n-doped case ($\mu>\Delta$), the Hall
conductivity for each valley has been obtained before as\cite{xiao2007,zhang2011,li2013}
\begin{equation}\label{sigmaH}
\sigma^0 _\text{H}
= -\tau _z \frac{{e^2 }}{2h}\left[ {1 - \frac{\Delta
}{{\mu }} - \frac{\Delta (\mu ^2-\Delta^2)}{{\mu ^3 }}} \right].
\end{equation}
Notice that the appearance of the valley index $\tau_z$ indicates
that the Hall conductivity also takes different signs between the two valleys.
In the absence of a magnetic field, the net effect is a pure valley Hall current
with a vanishing charge Hall current.

When a magnetic field is turned on,
from our discussion in the last section, there will be a field induced valley polarization. Therefore, the Hall current from
the two valleys cannot completely cancel each other and a net charge
current would appear.
In the low field regime, we can calculate the
charge Hall conductivity as
\begin{equation}\label{sigmac0}
\sigma^{0;c}_\text{H}= -\frac{{e^2 }}{h}\frac{\Delta}{ \mu
^2}\left( 1-\frac{3\Delta^2}{2 \mu^2} \right)\delta \mu,
\end{equation}
where $\delta \mu$ ($\ll \mu$) is the energy shift
due to the magnetic field (the difference between the shifted band bottoms).
From our discussion in section II.A,
\begin{equation}\label{deltamu}
\delta\mu=2mB\approx \frac{e\hbar v^2}{\Delta}B.
\end{equation}

\begin{figure}
\includegraphics[scale=0.9]{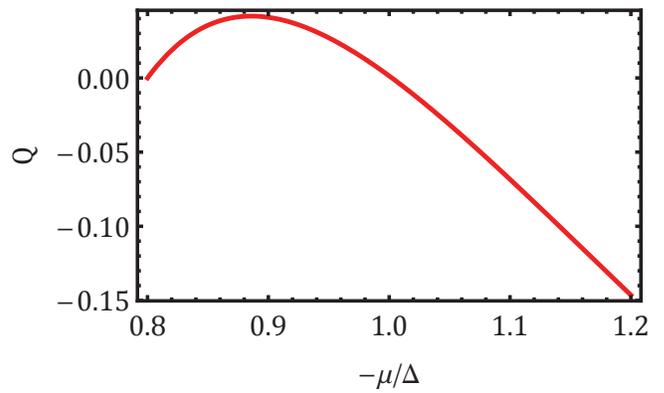}
\caption{(Color online). The variation of the dimensionless quantity $Q$ [see Eq.~\eqref{Eq:Qfactor}] with the chemical potential
$\mu$. Here, we take the spin-orbit coupling strength to be $\lambda=0.1\Delta$, and $v=0.5\times 10^6$\,m/s. \label{Fig:Qratio}}
\end{figure}

Let's focus on the Hall coefficient which is
an experimentally relevant physical quantity. It is defined as $\gamma=-\rho_\text{H}/B$.
The Hall resistivity $\rho_\text{H}$ is usually much less than the longitudinal
resistivity $\rho$. Hence we can write $\rho_\text{H}=\rho^2\sigma_\text{H}$.
Apart from the ordinary Hall coefficient
from the Lorentz force $\gamma_\text{OH}$,
now we have an extra contribution $\delta\gamma^0$ from the valley polarization.
From Eq.(\ref{sigmac0}) and (\ref{deltamu}), we have
\begin{equation}
\delta\gamma^0(\mu;\Delta)  = \frac{1}{e\mu D(\mu)}\left(1-
\frac{3\Delta^2}{2\mu^2} \right)\left( {\frac{{e^2 }}{h}\rho
 } \right)^2,
\end{equation}
where $D(\mu)=|\mu|/(2\pi \hbar^2 v^2)$ is the density of states for a single valley at energy $\mu$.
Note that this extra contribution has a sign change at $\mu=\sqrt{3/2}\Delta$, which can be
traced back to Eq.(\ref{sigmaH}) due to the different $\mu$ dependence of the intrinsic contribution and
the side jump contribution.

Now we consider the case of TMD. Because it consists of two copies of the massive Dirac model Eq.(\ref{H0}),
for the n-doped case ($\mu>\Delta$), the correction $\delta\gamma$ is simply a sum of
the contributions from both copies,
\begin{equation}
\delta\gamma=\delta\gamma^0(\mu-\lambda;\Delta-\lambda)+\delta\gamma^0(\mu+\lambda;\Delta+\lambda).
\end{equation}

The situation is simpler for the weakly p-doped case ($-\Delta+2\lambda>\mu>-\Delta-2\lambda$),
where only one copy with flavor $\tau_zs_z=1$ contributes. The result is given by
\begin{equation}
\delta\gamma=-\delta\gamma^0(-\mu+\lambda;\Delta-\lambda).
\end{equation}
In this case, the hole density can be estimated as
\begin{equation}
n_h\simeq 2\int_{\mu}^{-\Delta+2\lambda}D(\varepsilon)\text{d}\varepsilon=
\frac{1}{2\pi \hbar^2 v^2}(\mu-\Delta)(\mu+\Delta-2\lambda),
\end{equation}
where the factor of 2 appears because there are two valleys. The ordinary Hall
coefficient is given by $\gamma_\text{OH}=\frac{1}{n_h e}$. For comparison
we plot the dimensionless quantity
\begin{equation}\label{Eq:Qfactor}
\begin{split}
Q&=\left.\frac{\delta\gamma}{\gamma_\text{OH}}\right/\left( {\frac{{e^2 }}{h}\rho
}\right)^2\\
&=-\frac{(\mu-\Delta)(\mu+\Delta-2\lambda)}{(\mu-\lambda)^2}
\left[1-\frac{3(\Delta-\lambda)^2}{2(\mu-\lambda)^2}\right],
\end{split}
\end{equation}
in Fig.~\ref{Fig:Qratio} as a function of the chemical potential. The sign change
we noted before can be observed. Because $Q$ has
a magnitude on the order of $\sim (\lambda/\Delta)$ for low
doping levels ($\mu \sim -\Delta$), the correction from the valley polarization
would be more important with a decreasing gap $\Delta$ and an increasing spin
splitting. And it should be more pronounced for dirty samples with a large resistivity.

\section{discussion and summary}

From the above discussions, we see that due to the valley contrasting
orbital magnetic moments and Berry curvatures, the carriers at the two valleys
respond differently to the external fields.
In particular, the extra contribution to the Hall transport
is conceptually similar to what's happening in the anomalous Hall effect
in a ferromagnet.\cite{sini2008,naga2010}
In that case the carriers have a net spin
polarization and the the transverse motion is induced by the spin-orbit coupling.
In comparison, the valley here plays a similar role as the spin.
A population imbalance
between the two valleys is induced by the external magnetic field.
We can then ask: is it possible to construct an effective coupling between the valley
and the orbit motion that mimics the spin-orbit coupling for the Hall transport?

On a phenomenological level, there is indeed a
valley-orbit coupling, because the electrons in two valleys do have
opposite transverse velocities due to the opposite Berry curvatures.
The analogy with spin-orbit coupling can in fact be made more precise by a
systematic and rigorous procedure for deriving an effective single band
Hamiltonian.\cite{chan2008}
There are three basic ingredients for this procedure: the band energy, the
magnetic moment, and the Berry curvature.
The valley explicitly comes into the effective Hamiltonian at two places.
The magnetic moment carries the valley index and shifts the band energy
in a magnetic field by a Zeeman-like coupling term. There is also a dipole-like term proportional
to the electric field: $e\bm{E} \cdot \bm{R}$, where $\bm
R$ is the Berry connection which represents a shift of the wave packet center.\cite{chan2008}
For the conduction band of the simple model Eq.(\ref{H0}), it is
\begin{equation}
\bm{R}=-\frac{\tau_z}{2k^2}\left( {1 - \frac{\Delta }{{\sqrt
{\Delta ^2 + \hbar^2 v^2k^2  } }}} \right)\bm{k} \times
{\hat{\bm z}}.
\end{equation}
Therefore, the low-energy effective Hamiltonian
near the conduction band edge can be written as
\begin{equation}
H_\text{eff}  = \varepsilon (\bm{k}) -\tau_z\mu^*_B(\bm k)\;
\bm{B} \cdot \hat{\bm{z}}-e\phi({\bm r}) - \tau_z\frac{
e\hbar^{2}}{4m_e^{*2}v^2}  \hat{\bm{z}} \cdot
(\bm{k}\times \bm{E}),
\end{equation}
where we denote the magnitude of $\bm m(\bm k)$ as an effective Bohr magneton $\mu^*_B(\bm k)$.
We can see that the last term has a similar form as the
spin-orbit coupling, which is just the valley-orbit coupling. The effective Hamiltonian closely resembles
the Pauli Hamiltonian for an electron in the non-relativistic
limit derived from the Dirac equation. Taking this effective Hamiltonian as the starting point,
all the physical effects we discussed above could be addressed.

In summary, we have investigated the possibility of controlling the
valley degree of freedom using an applied magnetic field.
The valley-dependent orbital magnetic moment provides an essential
ingredient which couples the valley index with the magnetic field. It gives a nice
explanation for the asymmetric Landau level structure~\cite{li2013} and the enhanced magnetic susceptibility
that are common in systems with multiple massive Dirac fermions.

We point out that by tuning the external magnetic field
and adjust the doping level, we can efficiently
control the valley polarization. We also predict an extra contribution to the
ordinary Hall effect due
to the field induced valley-polarization. 

For pristine transition metal dichalcogenides material MoS$_2$, the typical values of 
model parameters are $v\sim 0.53\times 10^6$m/s, $\Delta\sim 0.83$eV, and $\lambda\sim 0.04$eV.\cite{xiao2012} 
The sudden change of magnetic susceptibility at the band edge on the order of $10^{-9}$A/T
should be detected. The two-step structure shown in Fig.~\ref{Fig:MagneticSusceptibillityTMD} is a quite
unique feature of this system. The spacing between the two steps in the valence band
corresponds to the spin-orbit splitting $4\lambda$. Hence this feature might be more clearly observed
in WS$_2$ and WSe$_2$ which have a larger spin-orbit coupling strength, $\lambda\sim 0.1$eV. 
The magnetic field induced valley polarization could be detected through circular dichroism
in optical transitions.\cite{zeng2012,mak2012,cao2012} For the weakly hole-doped case, it can also be detected by measuring the spin
polarization of the carriers because spin and valley are coupled in this case. 
As for both the field induced valley polarization and the resulting additional contribution to the Hall
transport, they increase as $\Delta$ decreases because both the Berry curvature and the orbital magnetic
moment scale as $1/\Delta$ near the band edge. Therefore, these effects could be enhanced
if the gap can be made smaller, possibly through chemical doping, straining or electrical gating
the two S or Se layers. 

Our results presented here are particularly
relevant for the valleytronics applications and for the study of 2D transition
metal dichalcogenides materials.
\\
\\
{\it Acknowledgments.} The authors would like to thank D. Goldhaber-Gordon and D.L. Deng
for valuable discussions and inspirations. S.A.Y. and X.L. are supported by
SUTD-SRG-EPD-2013062, the MOST Project of China (2012CB921300), and NSFC (91121004). T.C.
is supported by NSFC with Grant No. 11104193.
F.Z. is supported by DARPA (SPAWAR N66001-11-1-4110). W.Y. is supported by the HKSAR Research Grant Council with Grant No. HKU706412P and the Croucher 382 Foundation under the Croucher Innovation Award.
Q.N. is supported by Welch Foundation (F-1255) and DOE (DE-FG03-02ER45958, Division of Materials Science and Engineering).

\appendix*
\section{semiclassical quantization of Landau levels}

The energy of Landau levels can also be obtained by semiclassically
quantizing the cyclotron orbits.
The Onsager quantization condition states that the areas enclosed by the cyclotron orbits
should be quantized according to\cite{chan1996}
\begin{equation}\label{kkn}
\pi k_n^2 = \frac{{2\pi eB}}{\hbar }\left[n + \frac{1}{2} -
\frac{\Gamma(k_n) }{{2\pi }}\right],
\end{equation}
where $n=0,1,2,\cdots$ labels the orbits, and
$\Gamma(k_n)$ is the correction from the Berry phase
of the electron which is accumulated along the cyclotron orbit.
Practically, it can be calculated by integrating the Berry curvature
over the area enclosed by the orbit:
\begin{equation}
\Gamma(k_n)=\int_{S_n} \text{d}^2k\,\bm\Omega(\bm k)\cdot\hat{\bm z},
\end{equation}
After we obtain the quantized values of the wave-vector $k_n$ from Eq.(\ref{kkn}),
the Landau levels energies can be directly written down using the shifted band energy as
$\varepsilon_n=\varepsilon(k_n)-m(k_n)B$. We have checked that the Landau levels obtained using this semiclassical approach
agrees very well with the exact quantum result.


\begin{references}

\bibitem{ryce2007}     A. Rycerz, J. Tworzydlo, and C. W. J. Beenakker, Nat. Phys. \textbf{3}, 172 (2007).
\bibitem{xiao2007}     D. Xiao, W. Yao, and Q. Niu, Phys. Rev. Lett. \textbf{99}, 236809 (2007).
\bibitem{akhm2007}     A. R. Akhmerov and C. W. J. Beenakker, Phys. Rev. Lett. \textbf{98}, 157003 (2007).
\bibitem{yao2008}      W. Yao, D. Xiao, and Q. Niu, Phys. Rev. B \textbf{77}, 235406 (2008).
\bibitem{yao2009}      W. Yao, S. A. Yang, and Q. Niu, Phys. Rev. Lett. \textbf{102}, 096801 (2009).
\bibitem{zhang2011}    F. Zhang, J. Jung, G. A. Fiete, Q. Niu, and A. H. MacDonald, {Phys. Rev. Lett.} \textbf{106}, 156801 (2011).
\bibitem{zhu2012}      Z. Zhu, A. Collaudin, B. Fauque, W. Kang, and K. Behnia, Nat. Phys. \textbf{8}, 89 (2012).

\bibitem{novo2004}     K. S. Novoselov, A. K. Geim, S. V. Morozov, D. Jiang, Y. Zhang, S. V. Dubonos, I. V. Grigorieva, and A. A. Firsov,
    Science \textbf{306}, 666 (2004).
\bibitem{novo2005}     K. S. Novoselov, A. K. Geim, S. V. Morozov, D. Jiang, M. I. Katsnelson, I. V. Grigorieva, S. V. Dubonos, and A. A. Firsov, Nature \textbf{438}, 197 (2005).
\bibitem{zhan2005}     Y. Zhang, Y.-W. Tan, H. L. Stormer, and P. Kim, Nature \textbf{438}, 201 (2005).

\bibitem{aufr2010}     B. Aufray, A. Kara, S. Vizzini, H. Oughaddou, C. Le¡äandri, B. Ealet, and G. L. Lay, Appl. Phys. Lett. \textbf{96}, 183102 (2010).
\bibitem{pado2010}     P. E. Padova, C. Quaresima, C. Ottaviani, P. M. Sheverdyaeva, P. Moras, C. Carbone, D. Topwal, B. Olivieri, A. Kara, H. Oughaddou, B. Aufray, and G. L. Lay, Appl. Phys. Lett. \textbf{96}, 261905 (2010).
\bibitem{lalm2010}     B. Lalmi, H. Oughaddou, H. Enriquez, A. Kara, S. Vizzini, B. Ealet, and B. Aufray, Appl. Phys. Lett. \textbf{97}, 223109 (2010).

\bibitem{novo2005a}    K. S. Novoselov, D. Jiang, F. Schedin, T. J. Booth, V. V. Khotkevich, S. V. Morozov, and A. K. Geim, Proc. Natl. Acad. Sci. U.S.A. \textbf{102}, 10451 (2005).

\bibitem{sple2010}     A. Splendiani, L. Sun, Y. Zhang, T. Li, J. Kim, C.-Y. Chim, G. Galli, and F. Wang, Nano Lett. \textbf{10}, 1271 (2010).
\bibitem{mak2010}      K. F. Mak, C. Lee, J. Hone, J. Shan, and T. F. Heinz, Phys. Rev. Lett. \textbf{105}, 136805 (2010).
\bibitem{korn2011}     T. Korn, S. Heydrich, M. Hirmer, J. Schmutzler, and C. Schller, Appl. Phys. Lett. \textbf{99}, 102109 (2011).
\bibitem{xiao2012}     D. Xiao, G.-B. Liu, W. Feng, X. Xu, and W. Yao, Phys. Rev. Lett. \textbf{108}, 196802 (2012).
\bibitem{wang2012} Q. H. Wang, K. Kalantar-Zadeh,	 A. Kis, J. N. Coleman, M. S. Strano, Nat. Nanotech. \textbf{7}, 699 (2012).
\bibitem{zeng2012}     H. Zeng, J. Dai, W. Yao, D. Xiao, and X. Cui, Nat. Nanotech. \textbf{7}, 490 (2012).
\bibitem{mak2012}      K. F. Mak, K. He, J. Shan, and T. F. Heinz, Nat. Nanotech. \textbf{7}, 494 (2012).
\bibitem{cao2012}      T. Cao, G. Wang, W. Han, H. Ye, C. Zhu, J. Shi, Q. Niu, P. Tan, E. Wang, B. Liu, and J. Feng, Nat. Comm. \textbf{3}, 887 (2012).
\bibitem{wu2013}  S. Wu, J. S. Ross, G. Liu, G. Aivazian, A. Jones, Z. Fei, W. Zhu, D. Xiao, W. Yao, D. Cobden, X. Xu,
Nat. Phys. \textbf{9}, 149 (2013).
\bibitem{jone2013}     A. Jones, H. Yu, N. Ghimire, S. Wu, G. Aivazian, J. Ross, B. Zhao, J. Yan, D. Mandrus, D. Xiao, W. Yao, and X. Xu, Nat. Nanotech., advanced online publication, DOI: 10.1038/nnano.2013.151.

\bibitem{mak2013}  K. F. Mak, K. He, C. Lee, G. H. Lee, J. Hone, T. F. Heinz, J. Shan, Nat. Mat. \textbf{12}, 207 (2013).
\bibitem{ross2013} J. S. Ross, S. Wu, H. Yu, N. Ghimire, A. Jones, G. Aivazian, J. Yan, D. Mandrus, D. Xiao, W. Yao, X. Xu, Nat. Comm. \textbf{4}, 1474 (2013).

\bibitem{gong2013} Z. R. Gong, G. B. Liu, H. Y. Yu, D. Xiao, X. D. Cui, X. Xu, W. Yao, Nat. Comm. \textbf{4}, 2053 (2013).

\bibitem{radi2011}     B. Radisavljevic, A. Radenovic, J. Brivio, V. Giacometti, and A. Kis, Nature Nanotech. \textbf{6}, 147 (2011).

\bibitem{ye2013}       A. T. Neal, H. Liu, J. Gu, and P. D. Ye, arXiv:1308.0633, to appear in Nano Lett. (2013).

\bibitem{Bolotin}      Private discussions with Kirill Bolotin.

\bibitem{li2013}       X. Li, F. Zhang, and Q. Niu, Phys. Rev. Lett. \textbf{110}, 066803 (2013).

\bibitem{chan1996}     M.-C. Chang and Q. Niu, Phys. Rev. B \textbf{53}, 7010 (1996).
\bibitem{sund2000}     G. Sundaram, Ph.D. thesis, The University of Texas at Austin, 2000.
\bibitem{xiao2010}     D. Xiao, M.-C. Chang, and Q. Niu, Rev. Mod. Phys. \textbf{82}, 1959 (2010).

\bibitem{shar2004}     S. G. Sharapov, V. P. Gusynin, and H. Beck, Phys. Rev. B \textbf{69}, 075104 (2004).

\bibitem{mccl1956}     J. W. McClure, Phys. Rev. \textbf{104}, 666 (1956).
\bibitem{ando2007}     T. Ando, Physica E \textbf{40}, 213 (2007).

\bibitem{note1} Note that in real systems,
there are also contributions from the core electrons and lower lying valence bands to the orbital susceptibility,
which is not captured in the effective model.
However, the sudden change at band edge should still be visible.

\bibitem{mard2010}     P. Marder, Condensed Matter Physics (Wiley, New Jersey, 2010).


\bibitem{xiao2005}     D. Xiao, J. Shi, and Q. Niu, Phys. Rev. Lett. \textbf{95}, 137204 (2005).


\bibitem{jung2002}     T. Jungwirth, Q. Niu, and A. H. MacDonald, Phys. Rev. Lett. \textbf{88}, 207208 (2002).
\bibitem{onod2002}     M. Onoda and N. Nagaosa, J. Phys. Soc. Jpn. \textbf{71}, 19 (2002).

\bibitem{berg1970}     L. Berger, Phys. Rev. B. \textbf{2}, 4559 (1970).

\bibitem{note2} For the Gaussian type disorder, the so-called skew scattering contribution vanishes.

\bibitem{sini2008}     N. A. Sinitsyn, J. Phys:Condens. Matter \textbf{20}, 023201 (2008).

\bibitem{naga2010}     N. Nagaosa, J. Sinova, S. Onoda, A. H. MacDonald, and N. P. Ong, Rev. Mod. Phys. \textbf{82}, 1539 (2010).

\bibitem{chan2008}     M.-C. Chang and Q. Niu, J.Phys.:Condens. Matter \textbf{20}, 193202 (2008).



\end{references}
\end{document}